\documentclass[aps,prb,twocolumn,groupedaddress,byrevtex,showpacs,showkeys,amsfonts,amssymb,amsmath,floatfix]{revtex4-1}
\usepackage{graphicx}

\begin{document}

\title{Quasiparticle scattering time in superconducting films: from dirty to clean limit}

\author{A. Leo}
\email[]{antoleo@sa.infn.it}
\author{G. Grimaldi}
\author{R. Citro}
\author{A. Nigro}
\author{S. Pace}
\affiliation{CNR-SPIN and Dipartimento di Fisica 'E R Caianiello', Universit\`a di Salerno, Via Ponte Don Melillo, I-84084, Fisciano (SA), Italy}
\author{R.P. Huebener}
\affiliation{Physikalisches Institut, Universitaet Tuebingen, D-72076 Tuebingen, Germany}

\date{\today}

\begin{abstract}
We study the quasiparticle energy relaxation processes in superconducting Nb films of different thicknesses corresponding to different electron mean free paths in a state far from equilibrium, that is the highly dissipative flux-flow state driven up to the instability point. From the measured current-voltage curves we derive the vortex critical velocity $v^{*}$ for several temperatures. From the $v^{*}(T)$ values, the quasiparticle energy relaxation time $\tau_{\epsilon}$ is evaluated within the Larkin-Ovchinnikov model and numerical calculations of the quasiparticle energy relaxation rates are carried out to support the experimental findings. Besides the expected constant behavior of $\tau_{\epsilon}(T)$ for the dirty samples, we observe a strong temperature dependence of the quasiparticle energy relaxation time in the clean samples. This feature is associated with the increasing contribution from the electron-phonon scattering process.
\end{abstract}


\maketitle

\section{\label{sec:Intro}Introduction}
The quasiparticle energy relaxation time is one of the fundamental properties of superconductors which are essentially non-equilibrium in nature. Despite the great interest towards the comprehension of such mechanism, there are still few experimental techniques available to measure quasiparticle lifetimes and most of these are performed when the superconductor is slightly far from equilibrium as for example the quasiparticle injection, \cite{Levine1968} the photon irradiation \cite{Hu1974} or the measure of the quasiparticle-lifetime broadening in superconducting tunnel junctions. \cite{Dynes1978}\\
The study of the electronic flux-flow instability provides a way to extract information about the quasiparticle energy relaxation from an intrinsic bulk phenomena of the material when the quasiparticle distribution is far from equilibrium. \cite{Doettinger1997, Peroz-Villard2005, Kunchur2010} Indeed, the electronic flux-flow instability is a phenomenon associated with high velocity vortex motion, a dynamic regime characterized by a continuous diffusion of quasiparticles from the vortex core to the surrounding superconducting medium and by a re-trapping of quasiparticles into the vortex core. \cite{Klein1985}\\
Talking about the mechanisms which lead to the electronic flux-flow instability, we have to distinguish two limits: the high-temperature limit ($T \approx T_{c}$) and the low-temperature limit ($T \ll T_{c}$). Close to the critical temperature the Larkin-Ovchinnikov (LO) theory can be applied, \cite{LO} where the instability is associated to a shrinking of the moving vortex and to a drastic drop in the viscosity of superconducting medium due to the diffusion of quasiparticles from the vortex core to the surrounding. \cite{HuebenerBook} On the contrary, in the low-temperature limit there are three effects, i.e. the raising of the electronic temperature, the creation of additional quasiparticles and the diminishing of the superconducting gap $\Delta$, which lead to an expansion of the vortex. \cite{Peeters2007} This expansion has the effect to reduce the viscous drag because of the softening of gradients of the vortex profile. \cite{Kunchur2002}\\
Although the two temperature regimes are different, the flux-flow instability signature in the current-voltage characteristics is the same: an S-shape if voltage driven measurements are performed, or a sudden jump in voltage in the current driven ones. Moreover the instability parameters are directly related to the quasiparticle energy relaxation time $\tau_{\epsilon}$. In particular, in the LO framework it is:
	\begin{equation} \label{eq:tauvstar}
		\tau_{\epsilon} = \frac{D [14 \zeta(3)]^{1/2} (1 - \frac{T}{T_{c}})^{1/2}}{\pi v^{*2}}\,,
	\end{equation}
\noindent where $D$ is the quasiparticle diffusion coefficient, $\zeta(x)$ is the Riemann function and $v^{*}$ is the vortex critical velocity.\\
Finally, it is known that the quasiparticle energy relaxation process is ruled by different mechanisms in the dirty and clean limit. \cite{Peroz-Villard2005} In the dirty limit, the energy relaxation occurs within the vortex core, while in the opposite situation (clean limit) quasiparticles can relax the energy absorbed by the electric field only after experiencing a series of Andreev reflections and covering a total distance within the core equivalent to several coherence lengths $\xi$'s.\\
In this work we analyze the temperature dependence of the quasiparticle energy relaxation time as deduced by the flux-flow instability in a fixed magnetic field and in the range of temperature from $\approx 0.5 T_{c}$ to $\approx T_{c}$ on a set of Nb superconducting strips of increasing thickness to grasp the variation from the dirty to the clean limit. To support the experimental results a combined analysis based on the LO theory and the well known model of quasiparticle energy relaxation developed by Kaplan \emph{et al.} \cite{Kaplan1976} is carried out in the whole range of temperature to describe the crossover from dirty to clean limit. We observe that the temperature range of our experiment ($4.2 \div 10$ K) is well above the temperature value $T_{x} = r^{2} k_{B} T_{D}^{2} / \epsilon_{F}$, where $T_{D}$ is the Debye temperature, $\epsilon_{F}$ is the Fermi energy and $r$ is the phonon reflection coefficient at the film-substrate interface arising from acoustic mismatch and it is always $r < 1$. \cite{Kunchur2010} This temperature is the threshold above which the electron-electron scattering time is higher than the electron-phonon one and, consequently, the LO approach can be retained. We estimate from known parameters \cite{PooleBook} that for Nb this crossover temperature is $T_{x} = 1.1 r^{2}$~K.\\
We find that while the total quasiparticle relaxation time decreases toward the dirty limit, the electron-phonon scattering contribution increases with respect to the recombination one.

\section{\label{sec:model}Quasiparticle energy relaxation time}
Usually, quasiparticle energy relaxation rate $\tau_{\epsilon}^{-1}$ is considered as the result of the combination of different contributions from three different classes of processes: the electron-phonon scattering, the recombination to Cooper pairs and the inelastic part of electron-impurity scattering. The first kind of processes involves the changing of the quasiparticle excitation energy by means of interaction with phonon, while the second leads to the emission of a phonon by the two recombining particles to form bound Cooper pairs. \cite{Kaplan1976} In the limit of low concentration of non-magnetic impurities the electron-impurity scattering is usually neglected, indeed the properties of an s-wave superconductor are quite insensitive to this scattering process which plays a role only at very low temperature. \cite{Klapwijk2008}\\
The quasiparticles lifetimes due to the first and second type of processes described above are related to phonon density of states, $F(\Omega)$ ($\Omega$ being the phonon frequency), and the matrix element of the electron-phonon interaction, $\alpha^{2}(\Omega)$, within the strong coupling Eliashberg formulation \cite{Eliashberg1960, Scalapino1969} as discussed in Ref. \onlinecite{Kaplan1976}. Within this approach, the inverse of the lifetime, $\tau^{-1}_{\epsilon}(\omega)$ ($\omega$ being the quasiparticle frequency), is equal to twice the decay rate $\Gamma(\omega)$, which can be calculated by the single-particle Green's function \cite{Kaplan1976} or by means of the golden rule, i.e., by calculating the probability that the electron in the state with wavevector $\textbf{p}$ and frequency $\omega$ will emit a phonon with wavevector $\textbf{q}$ and frequency $\Omega$, and it's explicit expression is:
\begin{widetext}
	\begin{eqnarray} \label{eq:EQR3}
		\tau^{-1}_{\epsilon}(\omega) & = & \frac{2 \pi}{\hslash Z_{1}(0)} \int_{0}^{\omega - \Delta} d \Omega \alpha^{2}(\Omega) F(\Omega) \textrm{Re} \left( \frac{\omega-\Omega}{\sqrt{(\omega-\Omega)^{2} - \Delta^{2}}} \right) \left( 1 - \frac{\Delta^{2}}{\omega(\omega - \Omega)} \right) \left[ f(\Omega - \omega) + n(\Omega) \right] + \nonumber \\
		& + & \frac{2 \pi}{\hslash Z_{1}(0)} \int^{\infty}_{\omega + \Delta} d \Omega \alpha^{2}(\Omega) F(\Omega) \textrm{Re} \left( \frac{\omega-\Omega}{\sqrt{(\omega-\Omega)^{2} - \Delta^{2}}} \right) \left( 1 - \frac{\Delta^{2}}{\omega(\Omega - \omega)} \right) \left[ f(\Omega - \omega) + n(\Omega) \right] + \nonumber \\
		& + & \frac{2 \pi}{\hslash Z_{1}(0)} \int_{0}^{\infty} d \Omega \alpha^{2}(\Omega) F(\Omega) \textrm{Re} \left( \frac{\omega+\Omega}{\sqrt{(\omega+\Omega)^{2} - \Delta^{2}}} \right) \left( 1 - \frac{\Delta^{2}}{\omega(\Omega + \omega)} \right) \left[ f(\Omega - \omega) + n(\Omega) \right],
	\end{eqnarray}
\end{widetext}
where $f(\omega)$ and $n(\omega)$ are the Fermi and Bose distribution functions, respectively, while the spectral weight $\alpha^{2}(\Omega)F(\Omega)$ is given by averaging the square of the dressed electron-phonon matrix element over the Fermi surface, $Z_{1}(0)$ is the renormalization parameter and $\Delta$ is the superconducting gap.\\
The first and third term in Eq. \ref{eq:EQR3} represent the scattering processes with the emission and absorption of a phonon, respectively; they are denoted with $\tau_{s}$. The second term describes the recombination of two quasiparticles to form a pair and whose excess energy is emitted \text{as} a phonon; it is denoted with $\tau_{r}$. For $\omega \approx kT_{c}$ and $T \leq T_{c}$ only the low-frequency part of $\alpha^{2}F$ contributes where it behaves quadratically as $\approx b\Omega^{2}$, with $10^{3}b$ equal to $4~\textrm{meV}^{-2}$ for the Nb. \cite{Kaplan1976} The above expressions of $\tau_s$ and $\tau_r$ will be employed in the comparison with the quasiparticle relaxation times deduced from the vortex critical velocities.\\
Despite the fact that both electron-phonon scattering and recombination processes transfer directly the excess quasiparticle energy to the crystal lattice and (eventually) to the heat bath, the temperature dependence is distinctly different for both processes. Indeed, the scattering lifetime $\tau_{s}$ associated to the electron-phonon scattering increases as the temperature is lowered, with a power law of the type $\tau_{s} \propto T^{-n}$ with $n \approx 3$, owing to the decrease of the phonon population. At the gap edge $\omega = \Delta(T)$ the quasiparticle cannot emit a phonon and scatter because it is in the lowest energy state. For quasiparticles with energies $\omega > \Delta(T)$, spontaneous emission of a phonon sets a limit to the scattering lifetime $\tau_{s}$, thus $\tau_{s}$ becomes almost temperature independent. \cite{Kaplan1976}\\
The recombination lifetime, instead, must reflect the exponential temperature dependence of the quasiparticle population $e^{\Delta/k\,T}$. Another peculiarity is that for a quasiparticle at the gap edge $\tau_{r}$ goes through a minimum value for temperature close to $T_{c}$, while for $\omega > \Delta$ there is no minimum. \cite{Kaplan1976}\\
Let us note that when $T$ approaches $T_{c}$, the limiting values of $\tau_{s}$ and $\tau_{r}$ for $\omega = \Delta(T_{c}) = 0$ are equal due to particle-hole symmetry and their value is $\tau_{s(r)} (0, T_{c}) / \tau_{0} = 4.20$. Here $\tau_{0} = Z_{1}(0) \hslash / 2 \pi b(k T_{c})^{3}$ is a characteristic time which is used as a time unit to express $\tau_{s}$ and $\tau_{r}$ in an universal form. \cite{Kaplan1976}\\
Finally, we must consider that there are situations in which the branches of the quasiparticle excitation curve corresponding to quasiparticle wavevectors less than and greater than the Fermi wavevector are not equally occupied. In these situations, one can distinguish among the processes we are considering those which contribute to relaxation of the branch population imbalance. The associated lifetime is called the branch-mixing time $\tau_{Q}$. There is a relation between $\tau_{Q}$, $\tau_{r}$ and $\tau_{s}$: $\tau_{Q}^{-1} \leq \frac{1}{2} \left( \tau_{r}^{-1} + \tau_{s}^{-1} \right)$ (where the equality holds at the gap edge $\omega = \Delta$). \cite{Kaplan1976}

	\begin{table}[t]
	\caption{\label{tab:samples}Physical parameters of the Nb films.}
		\begin{ruledtabular}
			\begin{tabular}{clcccccc}
											& Sample 	& $w$ [$\mu$m] 	& $d$ [nm] 	& $T_{c}$ [K] 	& $l$ [nm]	& $\beta$\\
			\hline
			\emph{dirtiest}	& NbA02 	& 40 						& 30 				& 7.6 					& 1.6				& 1.7\\
			$\downarrow$ 		& NbA3 		& 50 						& 60 				& 8.1 					& 2.3 			& 2.1\\
			$\downarrow$ 		& NbC6 		& 50 						& 135 			& 8.7 					& 2.3 			& 2.3\\
			\emph{cleanest} & NbD8 		& 50 						& 150 			& 9.2 					& 3.7				& 2.6
			\end{tabular}
		\end{ruledtabular}
	\end{table}
\section{\label{sec:exp}Experimental details and results}
The samples investigated are Nb strips patterned by a standard UV photo-lithography technique on Nb films deposited on Si(100) substrates in a UHV dc diode magnetron sputtering system. The film thickness $d$ was controlled via a quartz crystal monitor calibrated by low-angle reflectivity measurements. The Nb strips width $w$ is $40~\mu\textrm{m}$ for the sample named NbA02 and $50~\mu\textrm{m}$ for samples NbA3, NbC6 and NbD8 \cite{Grimaldi2010_PRB}. The distance $L$ between the voltage tips for all the samples is 2 mm long. A typical value of the critical current density, obtained by the standard $10~\mu\textrm{V}/\textrm{cm}$ criterion, at 4.2 K and zero magnetic field is $J_c = 2 \cdot 10^{6}~\textrm{A}/\textrm{cm}^{2}$.\\
The characteristic parameters of the samples are summarized in Table \ref{tab:samples}. The electron mean free path $l$ is estimated via the relation $l = \frac{1}{v_{F} \gamma \rho_{N}} \left(\frac{\pi k_{B}}{e}\right)^{2}$, \cite{Hauser1964} where $v_{F} = 2.73 \cdot 10^{5}$~m/s is the Fermi velocity, \cite{Kerchner1981} $\gamma = 7 \cdot 10^{2}~\textrm{J}/\textrm{K}^{2} \textrm{m}^{3}$ is the Nb electronic specific heat coefficient \cite{ChemBook} and the resistivity $\rho_{N}$ is estimated at $T = 10$~K. The mean value of the residual resistance ratio is around $\beta = R(300K)/R(10K) \approx 2$, while it increases as the thickness increases (see Table \ref{tab:samples}, thus the samples are getting progressively clean with increasing thickness.\\
As it is well known, when $\xi_{0} \ll l$ we are dealing with the clean limit, while for $\xi_{0} \gg l$ the sample is in the dirty limit. Thus, comparing the values of $l$ in Table \ref{tab:samples} with the value $\xi_{0} \approx 7$~nm of the coherence length as obtained by $H_{c2}(T)$ measurements and assumed to be independent of $d$, \cite{Cirillo2005} we can argue that the whole set of samples covers the crossover region between the two limits. The reduction of the mean free path with decreasing thickness of the samples may be due to the poor crystalline quality of Nb film in proximity of the interface with the substrate.\\
	\begin{figure}[t]
	\begin{center}
	\includegraphics[width=0.45\textwidth]{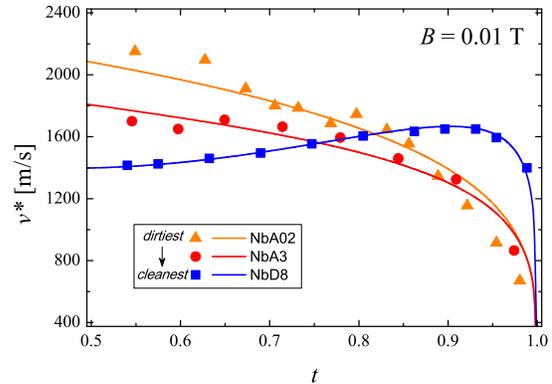}
	\caption{\label{fig:F1} Experimental $v^{*}(t)$ curves at $B = 0.01$ T of samples NbA02 (orange triangles), NbA3 (red circles) and NbD8 (blue squares). The solid lines are the resulting fitting curves obtained by Eq. \ref{eq:vstarT} assuming for $\tau_{\epsilon}$ the form in Eq. \ref{eq:taufit}, where $\tau_{s}$ and $\tau_{el}$ are the fitting parameters.}
	\end{center}
	\end{figure}
\noindent In order to analyze the mean free path influence on the quasiparticle energy relaxation time, we use as a tool the flux-flow instability phenomenon which is observed in the $V(I)$ curves of the samples for a wide temperature range. \cite{Grimaldi2009_PRB, Grimaldi2010_PRB} In particular, in Fig. \ref{fig:F1} are shown the $v^{*}(t)$ (with $t = T/T_{c}$) for the three samples of increasing $l$: NbA02, NbA3 and NbD8. The curves are extracted from current driven $V(I)$ measurements performed in a fixed magnetic field $B = 0.01$~T with a pulsed current 4-probe technique in order to minimize self-heating effects. \cite{Grimaldi2008_EuCAS} The $v^{*}$ values are evaluated via the formula $v^{*} = V^{*}/(B \, L)$, where $V^{*}$ is the voltage value at the instability point. For the dirty samples NbA3 and NbA02 the $v^{*}(t)$ shows a monotonous decrease of the vortex critical velocity as the temperature is raised, in agreement with the LO expectation. \cite{LO} On the contrary, the $v^{*}(t)$ for the cleanest sample NbD8 presents the evident existence of a maximum before the predicted decrease. This unusual behavior can be ascribed to the clean nature of the considered sample, as it is explained in Section IV.\\
Let us note that, even if the temperature range is not so close to $T_{c}$, the use of the LO approach is further sustained by the evaluation of the value of the Bezuglyj and Shklovskij parameter $B_{T}$. \cite{BS1992} Indeed, the Bezuglyj and Shklovskij (BS) approach extends the LO model taking into account quasi-particle heating due to the finite heat removal rate of power dissipated in the sample. In the BS approach a macroscopic parameter $B_{T}$ is derived from a microscopic analysis of the heat removal as a result of the mutual phonon exchange rather than interface properties between the film and the substrate \cite{BS1992}. The $B_{T}$ parameter separates the region where non thermal ($B \ll B_{T}$) or pure heating mechanisms ($B \gg B_{T}$) of the instability dominates. In our measurements $B_{T}$ at 4.2~K is 0.24~T, \cite{Grimaldi2008_EuCAS} i.e. a value well above the applied magnetic field $B = 0.01$~T. Thus, the quasiparticle overheating, which is effective only when the magnetic field value exceed $B_{T}$, is negligible.\\
Moreover, the magnetic field dependence of the dissipated power density $p^{*} = J^{*} \cdot E^{*}$ is considered (see Fig. \ref{fig:F2new}), where $J^{*}$ is the current density and $E^{*}$ is the electric field value at which the instability occurs. There is a monotonous increase of $p^{*}$ with the increasing field which indicates that the self-heating effects are negligible. \cite{Xiao1999}
	\begin{figure}[t]
	\begin{center}
	\includegraphics[width=0.45\textwidth]{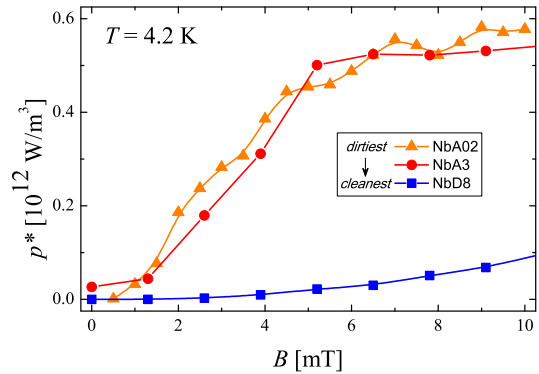}
	\caption{\label{fig:F2new} Dissipated power density $p^{*}$ as a function of the applied magnetic field $B$ at $T = 4.2K$ for the same three samples of Fig. \ref{fig:F1}. The solid lines are guide for the eyes.}
	\end{center}
	\end{figure}

\section{\label{sec:discussion}Discussion}
\subsection{On experimental data}
	\begin{table*}[t]
	\caption{\label{tab:fits}Estimated quasiparticle energy relaxation times.}
		\begin{ruledtabular}
			\begin{tabular}{clccccc}
											&Sample &$\tau_{el}$ [ns] &$\tau_{s}$ [ns] &$\tau_{\epsilon}(t=0.5)$ [ns]	&$\tau_{0}$ [ns]	&$p^{*}(t=0.5)$ [$10^{12}$~W/m$^3$]\\
			\hline
			\emph{dirtiest}	&NbA02 	&$\approx 5$ 			&$0.33 \pm 0.01$ &0.32													&$0.26$ 					&0.58\\
			$\downarrow$		&NbA3 	&$1.60 \pm 0.90$ 	&$0.42 \pm 0.01$ &0.42													&$0.22$ 					&0.54\\
			\emph{cleanest} &NbD8 	&$0.22 \pm 0.01$ 	&$0.71 \pm 0.01$ &0.66													&$0.15$ 					&0.35\\
			\end{tabular}
		\end{ruledtabular}
	\end{table*}
	\begin{figure}[b]
	\begin{center}
	\includegraphics[width=0.45\textwidth]{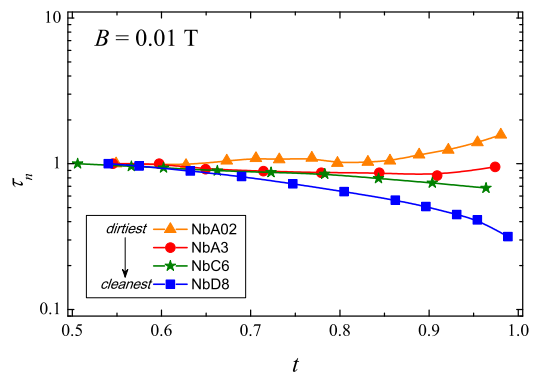}
	\caption{\label{fig:F2} Evaluated $\tau_{n}(t)$ curves by Eq. \ref{eq:tauvstar} at $B = 0.01$ T for samples NbA02 (orange triangles), NbA3 (red circles), NbC6 (green stars) and NbD8 (blue squares). Solid line are guide for eyes.}
	\end{center}
	\end{figure}
The experimental data presented in Fig. \ref{fig:F1} related to the dirty samples NbA02 and NbA3 fairly agree with the LO prediction on a monotonous decrease of the vortex critical velocity as the temperature approaches $T_{c}$. This behavior is associated with a temperature independent quasiparticle energy relaxation rate $\tau_{\epsilon}^{-1}$. \cite{LO} This prediction is confirmed by the evaluated $\tau_{n}(t) = \tau_{\epsilon}(t)/\tau_{\epsilon}(0.5)$ curves shown in Fig. \ref{fig:F2}, where the $\tau_{\epsilon}(t)$ are estimated from the experimental $v^{*}(t)$ curves inverting the relation:
	\begin{equation} \label{eq:vstarT}
		v^{*} = \left[ \frac{\left( 1 - t \right)^{1/2} \, D \, \left[14 \, \zeta(3)\right]^{1/2}}{\pi \, \tau_{\epsilon}} \right]^{\frac{1}{2}} \cdot \left( 1 + \frac{a_{0}}{\sqrt{D \, \tau_{\epsilon}}} \right)\,,
	\end{equation}
\noindent with $a_{0} = \left( \frac{2}{\sqrt{3}} \frac{\Phi_{0}}{B} \right)^{1/2}$ the inter-vortex distance and $D = \frac{1}{3} v_{F} l$. Eq. \ref{eq:vstarT} is a derived expression from the LO which takes into account the necessary condition of spatial homogeneity of the non-equilibrium quasiparticle distribution between the vortices in the range of low magnetic fields. \cite{Doettinger1995, Lefloch1999}\\
From Fig. \ref{fig:F2} it is clear that for all three dirty samples (namely, NbA02, NbA3 and NbC6) an almost temperature independent $\tau_{n}(t)$ is observed. The slight increase related to samples NbA02 and NbA3 for $t > 0.9$ will be discussed in the part B of this Section.\\
On the contrary, in Fig. \ref{fig:F1} a strong non-monotonic behavior of the $v^{*}(t)$ related to cleanest sample NbD8 can be clearly seen. This behavior is associated with a temperature dependent $\tau_{\epsilon}$ (see Fig. \ref{fig:F2}). Indeed, in this limit the contribution of the recombination processes is not negligible and it is known that these processes are strongly influenced by the temperature (see Section \ref{sec:model}).\\
In order to estimate the contribution of the recombination processes we performed a non-linear fitting procedure on the data shown in Fig. \ref{fig:F1} on the basis of Eq. \ref{eq:vstarT} by assuming the quasiparticle energy relaxation rate $\tau_{\epsilon}^{-1}$ as the sum of two contributions: one is a temperature independent term $\tau_{s}^{-1}$, which takes into account the scattering processes, and the other one is an exponentially temperature dependent term $\tau_{r}^{-1} = \tau_{el}^{-1}\textrm{Exp}\left[ -1.76\,m\,\frac{\sqrt{1-t}}{t} \right]$ which takes into account the recombination processes:
	\begin{equation} \label{eq:taufit}
		\frac{1}{\tau_{\epsilon}} = \frac{1}{\tau_{s}} + \frac{1}{\tau_{el}}\textrm{Exp}\left[ -1.76\,m\,\frac{\sqrt{1-t}}{t} \right]\,,
	\end{equation}
\noindent where $m$ is a numerical parameter. The assumption to consider the scattering contribution as independent from temperature follows from the observation that when the instability is triggered, the quasiparticle energy is much higher than the energy gap, as we demonstrate below.\\
In Table \ref{tab:fits} are summarized the results of the fitting procedure. The values of the two fitting parameters $\tau_{el}$ and $\tau_{s}$ were obtained by fixing the parameter $m$ equal to 1.5, which can be interpreted in the framework of recombination processing involving more than two quasiparticles. \cite{Doettinger1997} If we compare for each sample the $\tau_{s}$ and $\tau_{el}$ values, we recognize that for the dirty samples (namely NbA02 and NbA3) the faster process is the scattering one, thus it is the one which dominates the quasiparticle energy relaxation and it results $\tau_{\epsilon} \approx \tau_{s}$. On the contrary, for the cleanest sample (namely NbD8) it is the recombination process which dominates over the scattering one. As shown in Fig. \ref{fig:F1}, the exponential dependence on temperature of the recombination processes determines the peculiar non monotonic behavior of the vortex critical velocity in this sample. These results point out that the contribution from the recombination process becomes more significant as the sample becomes more clean.\\
Theoretical and experimental studies on the temperature dependence of the lifetime broadening factor $\Gamma$ of the tunneling conductance curves gave also evidence of an electron-phonon scattering rate larger than the recombination one in disordered superconductors. \cite{White1986, Devereaux1991, Pyun1991} Furthermore a power law dependence of the scattering rate on temperature ($\approx T^{3}$) has been deduced from these tunneling experiments, whereas an almost constant behavior of  $\tau_{s}$ can be inferred from our measurements. Nevertheless, in agreement with a previous study on the flux flow properties of clean and dirty samples, \cite{Peroz-Villard2005} here we find the experimental evidence of a crossover between two different behaviors: for dirty samples the electron-phonon scattering dominates, while for the clean samples the recombination to Cooper pairs plays the significant role, and the change in the balance between the two contributions is continuous.\\
Moving from the dirty to the clean limit $\tau_{el}$ decreases, while $\tau_{s}$ increases as well as the total quasiparticle energy relaxation time $\tau_{\epsilon}$ (evaluated via Eq. \ref{eq:taufit}), as it can be seen in Table \ref{tab:fits}. The decrease of the scattering time $\tau_{s}$ in the thinner samples and its temperature behavior may be ascribed to an additional relaxation channel due to the scattering by impurities and/or to the high energy quasiparticles involved in the conventional electron-phonon interaction. Indeed, whereas in first approximation the energy gap and the thermodynamic properties of superconductors remain unchanged in the presence of a low concentration of impurities, \cite{AbrikosovBook} our results may indicate that the situation is more complex and the inelastic part of the impurity scattering may play a role.\\
	\begin{figure}[t]
	\begin{center}
	\includegraphics[width=0.45\textwidth]{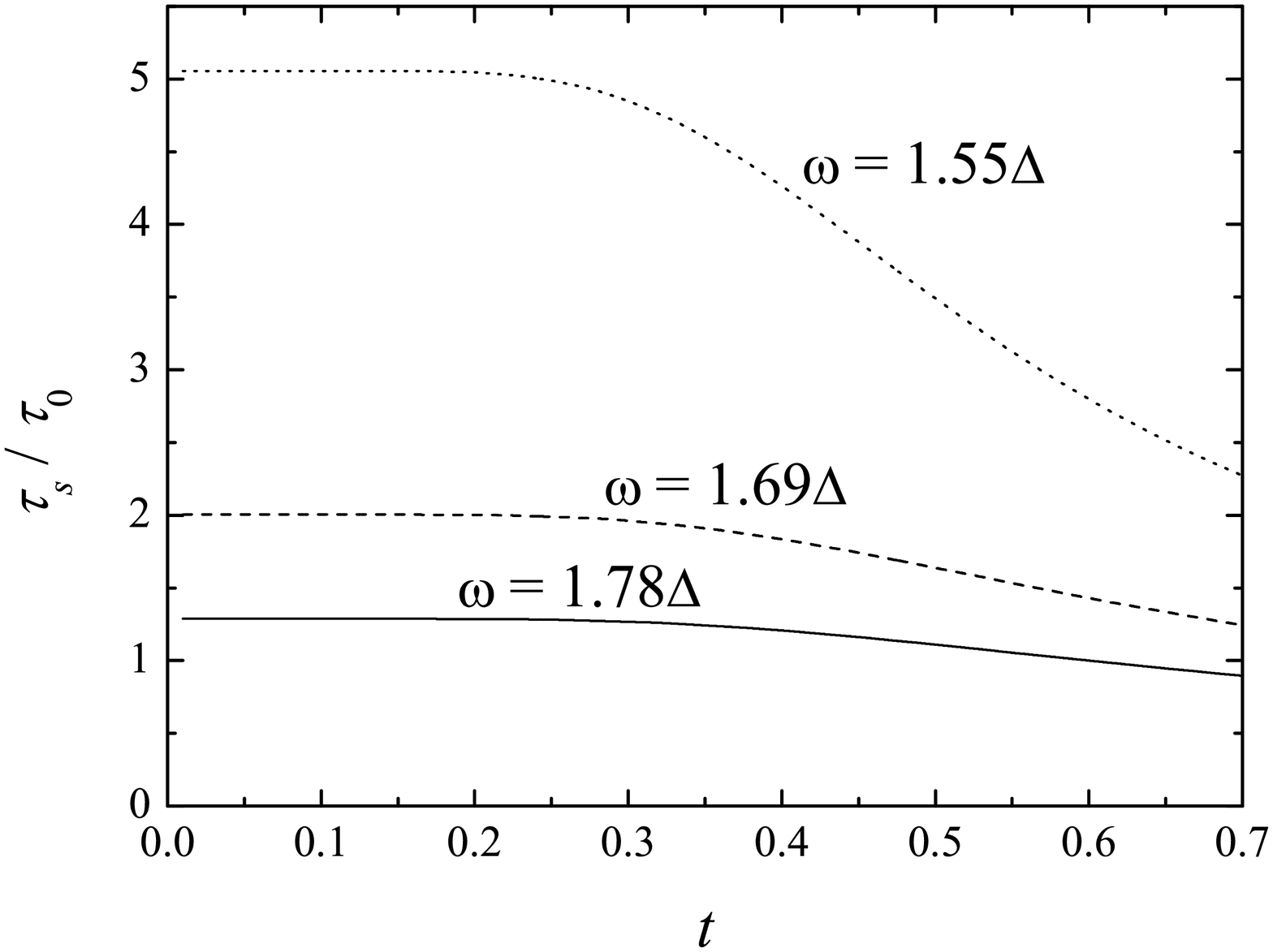}
	\caption{\label{fig:F3} Scattering time $\tau_{s}$ in units of $\tau_{0}$ as a function of the reduced temperature $t = T / T_{c}$ for different values of the quasiparticle energy: $\omega = 1.55\Delta$ (dotted line), $1.69\Delta$ (dashed line) and $1.78\Delta$ (solid line).}
	\end{center}
	\end{figure}
On the other hand, in the limit of high quasiparticle excitation energy a drastic decrease and a nearly constant temperature dependence of the electron-phonon scattering lifetime is predicted. \cite{Kaplan1976} In Fig. \ref{fig:F3} there is a plot of the scattering time $\tau_s(t) / \tau_0$ for different values of the ratio $\tau_{s}(t \rightarrow 0)/\tau_{0}$ (with $\tau_{s}(t \rightarrow 0)$ assumed to be almost equal to the estimated $\tau_{s}$ in Table \ref{tab:fits}), which correspond to different quasiparticle energies above the gap. As shown, the more the sample is dirty, the more the scattering time is lowered.\\
To support this scenario, an indication of the energy scale of quasiparticle excitations involved at the vortex instability point comes from the density of electric power $p^{*}$ for the three investigated samples (see Table \ref{tab:fits}). The $p^{*}$ values increase going towards the dirty sample and this is indicative of the presence of an high energy quasiparticle population.\\
	\begin{figure}
	\begin{center}
	\includegraphics[width=0.45\textwidth]{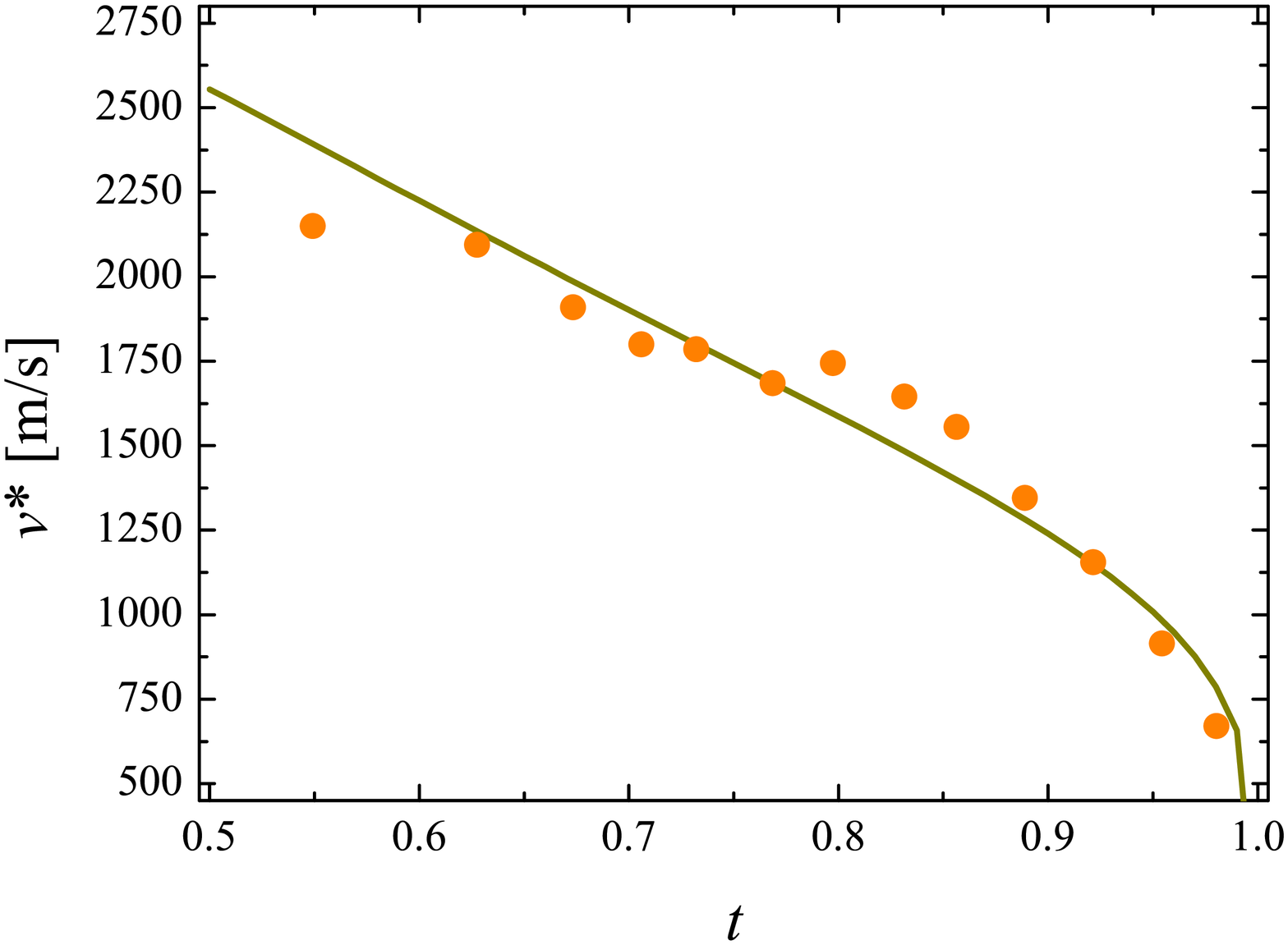}
	\caption{\label{fig:F4} Comparison between the experimental $v^{*}(t)$ curve of sample NbA02 (dots) and the one obtained by Eq. \ref{eq:vstarT} with a value of $\tau_{s}$ corresponding to quasiparticles with excitation energy $\omega = 1.78 \Delta$ (solid line).}
	\end{center}
	\end{figure}
\noindent This is further confirmed by comparing the experimental data of $v^{*}(t)$ with the curve obtained by Eq. \ref{eq:vstarT} using the expression of the scattering time $\tau_{s}$ resulting from Eq. \ref{eq:EQR3}, as shown in Fig. \ref{fig:F4}. To compare the two curves, one has to determine the quasiparticle energy (above the gap) by the knowledge of the ratio $\tau_{s}(t \rightarrow 0)/\tau_{0}$. In particular the solid line in Fig. \ref{fig:F4} includes a scattering lifetime $\tau_{s}$ for quasiparticles with excitation energy $\omega = 1.78 \Delta$, well above the gap edge.\\
Thus, it appears that the electron-phonon scattering contribution in dirty samples prevails on the recombination one, and its weaker temperature behavior deduced from our data seems to be related to the quasiparticles from different energy scales involved in the electronic flux-flow instability mechanism.

\subsection{Other contributions to the quasiparticles scattering rates: the branch mixing}
\begin{figure}
	\begin{center}
	\includegraphics[width=0.45\textwidth]{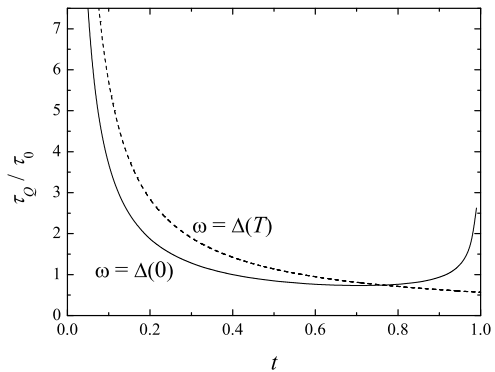}
	\caption{\label{fig:F5} Branch-mixing time $\tau_{Q}$ in units of $\tau_{0}$ as a function of the reduced temperature $t = T / T_{c}$ for $\omega = \Delta(0)$ (solid line) and $\omega = \Delta(T)$ (dashed line).}
	\end{center}
	\end{figure}

As discussed previously, the observed $\tau_{\epsilon}(t)$ behavior is in good agreement for the dirty and clean samples with literature. \cite{Peroz-Villard2005, AngrisaniArmenio2007} However, there is a slight discrepancy in the $\tau_{\epsilon}(t)$ curves of the dirty samples NbA02 and NbA3, which show a clear up bended curvature as the temperature approaches $T_{c}$ (see Fig. \ref{fig:F2}). This can be interpreted as a signature of the presence of additional scattering processes associated to a highly unbalanced distribution of quasiparticles in the levels below and above the Fermi level, the so called branch-mixing process.\\
In fact, while in thermal equilibrium the branches of quasiparticles with wavevector below and above the Fermi surface are equally populated, in the presence of a current flow across an interface or due to the injection of quasiparticles through tunnel barriers, a branch population imbalance can occur. The contribution to the quasiparticle decay rate due to the branch mixing can be calculated by the golden rule or by keeping extra terms in the Nambu self-energy. \cite{Kaplan1976} In particular, when the quadratic approximation is used for the factor $\alpha(\Omega)F(\Omega)$ and the BCS relation $2\Delta = 3.52 k T_{c}$ is considered, a useful approximate form of $\tau_{Q}$ for $\omega \gg \Delta(T)$ is:\\
	\begin{equation} \label{eq:EQR6}
		\frac{\tau_{0}}{\tau_{Q}(\omega, T)} \approx \frac{\Delta(T)}{2\Delta(0)} \left( \frac{\omega}{k T_{c}} \right)^{2} \left( \frac{\Delta(0)}{k T_{c}} \right) \textrm{coth}\left(\frac{\omega}{2 k T}\right).
	\end{equation}
As shown $\tau_{Q}$ diverges as $\Delta^{-1}(T)$ for $T \rightarrow T_{c}$. For other ranges of the excitation energy $\omega$ the following relation holds, $\tau_{Q}^{-1} \leq \frac{1}{2} (\tau_{s}^{-1} +\tau_{r}^{-1}$), where the equality holds at the gap edge $\omega = \Delta(\omega)$. This leads to a limiting value of $\tau_{Q}^{-1}(\Delta(T))$ at $T = T_{c}$ given by $0.73$. A plot of $\tau_{Q}$ is reported in Fig. \ref{fig:F5} for $\omega = \Delta(T)$ and $\omega = \Delta(0)$. One notices that at the band edge a minimum develops close to $T_{c}$.\\
We can suppose that the mechanism responsible for the unbalanced distribution of quasiparticles in the levels below and above the Fermi level leading to the branch mixing is the same mechanism responsible for the instability in the dirty samples.

\section{Conclusions}
Summarizing, we have investigated the quasiparticle energy relaxation processes from dirty to clean superconducting films by studying the temperature dependence of the electronic flux-flow instabilities. We have found that the vortex critical velocity behavior as a function of the temperature $v^{*}(t)$ and the resulting quasiparticle energy relaxation time $\tau_{\epsilon}(t)$ are strongly modified when the sample changes from the dirty to the clean limit. From our findings we can argue that the vortex instability mechanism involves different quasiparticle energy scales in Nb films with different electron mean free paths, giving rise to a more important contribution of the electron-phonon scattering on the energy relaxation time in dirty samples. It appears that the fraction of the quasiparticles experiencing scattering is eliminated from the pairing process.\\
These conclusions are supported by numerical calculation based on the model developed by Kaplan \emph{et al.} for the quasiparticle energy relaxation processes. \cite{Kaplan1976} Indeed, the simulations show how the $v^{*}(t)$ related to the dirty samples can be reconstructed considering only the contribution from the quasiparticle scattering, while for the clean limit it is necessary to consider also the contribution by the quasiparticle recombination.\\
Moreover, our investigation enlightens a deviation of $\tau_{\epsilon}(t)$ from the usually constant behavior of the extreme dirty samples. Indeed, the $\tau_{\epsilon}(t)$ curves show a bend up curvature for $T \approx T_{c}$. This unexpected behavior is associated to a non-negligible contribution to the quasiparticle energy relaxation from the branch-mixing processes. It is supposed that the cause of the unbalancing distribution of quasiparticles in the levels close to the Fermi level is the instability mechanism.\\
Recently, we became aware of overlapping, simultaneous work, \cite{Kunchur2010} in which the energy relaxation time in low pinning molybdenum-germanium films has been measured by the vortex instability tool. We hope that further experimental and theoretical work will be carried out in order to deeper clarify this subject.

\begin{acknowledgments}
We are grateful to N. Schopohl for useful discussions. We would thank C. Attanasio and C. Cirillo for providing Nb thin films. This work was partially supported by the Research Project L.R. N°5, Regione Campania. A.L. acknowledges the funding support under the Contract No. CO 03/2009.
\end{acknowledgments}


\end{document}